\documentstyle[preprint,prl,aps]{revtex}
\tightenlines
\begin{document}
\preprint{TNT 97}
\draft

\title{Characterization of the spatial complex 
      behavior and transition to chaos in flow systems}

\author{M. Falcioni $^{(\ast)(a)}$, D. Vergni 
        and A. Vulpiani $^{(a)(b)}$}

\address{Dipartimento di Fisica, Universit\`a ``La Sapienza'',
         P.le A. Moro 2, 00185, Roma, Italy}
\address{($\ast$) Corresponding author (fax:+39-6-4463158;
     E-mail: M.Falcioni@roma1.infn.it)}
\address{(a) also INFN, Sezione di Roma 1}
\address{(b) also INFM, Unit\`a di Roma 1}

\date{\today}

\maketitle

\begin{abstract}
We introduce a  ``spatial'' Lyapunov exponent to characterize 
the complex behavior of non chaotic but convectively unstable 
flow systems. This complexity is of spatial type and is due to 
sensitivity to the boundary conditions. We show that there 
exists a relation between the spatial-complexity index we 
define and the comoving Lyapunov exponents. In these systems 
the transition to chaos, i.e. the appearing of a positive 
Lyapunov exponent, can take place in two different ways. In 
the first one (from neither chaotic nor spatially complex 
behavior to chaos) one has the typical scenario; that is, 
as the system size grows up the spectrum of the Lyapunov
exponents gives rise to a density. In the second one (when the 
chaos develops from a convectively unstable situation) one 
observes only a finite number of positive Lyapunov exponents.

\end{abstract}



\section{Introduction}
The dynamical chaos is considered to be one of the main sources
of complex behavior in a dynamical system. One of the main 
properties of dynamical chaos is the sensitive dependence of the  
evolution on the initial conditions, i.e.: a small error
on the initial state grows exponentially in time  \cite{eck}. This 
behavior is usually assumed as the characterizing property 
of chaos, and it is quantified by a positive value of the 
maximal Lyapunov exponent $\lambda _1$. 

However, highly nontrivial behaviors 
can appear also in systems which are not chaotic (i.e., 
$\lambda _1 \leq 0$). Let us mention the systems with 
asymptotically stable fixed points, but with fractal 
boundaries of the attraction basins \cite{ott}, and the 
chaotic scattering phenomenon \cite{chasca}, where the 
``chaos'' is just transient. 

An interesting situation can occur in high dimensional 
systems, like the following chain of maps with unidirectional 
coupling: 
\begin{equation}
\label{eq:uno}
x_n(t+1)=(1-c)f_a(x_n(t))+ c f_a(x_{n-1}(t)),
\end{equation}
where $t$ is the discrete time, $n=1,2,3,\dots, N$ is a 
spatial index and $x_0(t)$ is a given boundary condition.
These models are quite natural candidates for the 
description of flow systems, that are systems with a 
privileged direction, e.g.: boundary layer, thermal convection and 
wind-induced water waves \cite{afra}.  

After the seminal papers of Deissler and Kaneko \cite{deisk} 
it is now well known that nontrivial phenomena can take 
place in systems with asymmetric couplings, even in the absence 
of chaos ($\lambda _1 \leq 0$). In particular, if the system is 
convectively unstable the spatial structure can be very complex 
and the external noise can have an important role in the formation 
and the maintenance of the structure \cite{deisk,deis}. 
In spite of the clear evidence of a spatial ``complexity''
in these non chaotic systems, up to now, as far as we know, 
there is not a simple and systematic quantitative 
characterization of this phenomenon. To answer this purpose, 
we define a quantity that measures the degree of sensitivity 
of the system to the boundary conditions, and we study its possible  
quantitative relation with the comoving Lyapunov exponents
-- the quantities by means of which one can define the convective 
instabilities. The definition of the comoving Lyapunov exponents 
$\lambda(v)$, for these extended systems, may be given as follows 
\cite{deisk}. If $\delta x_0 (0)$ is a perturbation on the 
boundary at the time $t=0$, in a frame of reference that moves 
along the system with velocity $v > 0$, at large $t$, this 
perturbation is $O(\delta x_0 (0) \exp [\lambda(v)t])$. 
If $\lambda(v) < 0$ for all $v>0$ the system is said to be 
absolutely stable; if there exists a range of velocity 
for which $\lambda(v)$ is positive, then the system is said to 
be convectively unstable. The interesting situation, in a 
convectively unstable system, arises when the usual Lyapunov 
exponent, $\lambda_1 =\lambda(v=0)$, is negative. 

In sect. II we discuss some results about the flow system 
model (\ref{eq:uno}) with $f_a(x)=a\, x\, (1-x)$, that is 
the local map we use for all the computations. In particular 
we report on the qualitative spatio-temporal behaviors 
at varying the control parameters $c$ and $a$. 

In sect. III we introduce an index -- we call it ``spatial 
Lyapunov exponent''-- that supplies us with a quantitative 
characterization of the spatial complexity, in terms of 
the spatial sensitivity to the boundary conditions. We show 
that there exists a strong relation between the  ``spatial'' 
Lyapunov exponent and the comoving Lyapunov exponents. 

In sect. IV the reader can find a study on the transition 
to chaos for the flow systems. The transition to 
a positive value of $\lambda _1$ 
can take place in two different settings:   
\begin{description} 
\item[I)] from an absolutely stable state (i.e., a state 
for which the comoving Lyapunov exponents are negative for 
all velocities), that is not spatially complex;  
\item[II)] from a state that is convectively unstable 
(i.e., a state for which the comoving Lyapunov exponents 
are positive for some velocities), that has a certain degree 
of spatial complexity.
\end{description}
The case ${\bf I})$ is a rather standard transition, 
by which we mean that, for large $N$, there exists 
a finite density of positive Lyapunov 
exponents. On the other hand, in the case ${\bf II})$, 
at varying $N$, one obtains only a finite number, not 
a finite fraction, of positive Lyapunov exponents. The latter 
behavior is a clear indication that the transition is 
dominated by a sort of boundary effect. 

Sect. V is devoted to conclusions and discussion.

\section{Qualitative behavior of the model}
In this section we present some dynamical features of the 
unidirectionally coupled map lattice (\ref{eq:uno}), with 
$f_a(x)=a\, x\, (1-x)$. In many papers the boundary condition 
is kept fixed, i.e. $x_0 (t)=x^{\ast}$, and often $x^{\ast}$ 
is an unstable fixed point of the single map $x(t+1)=f_a (x(t))$ 
\cite{fix}. Here, following Deissler \cite{deisk} and Pikovsky 
\cite{piko}, we adopt a more general time dependent boundary 
condition: $x_0 (t)=y(t)$ with $y(t)$ a known function, that 
may be periodic, quasi-periodic or obtained by a chaotic system. 

At varying the control 
parameters, $c$ and $a$, one observes a plethora of different 
spatio-temporal behaviors. In Fig. \ref{fig:4_16} we show 
the results of a numerical exploration of the phase space 
of the system, with a quasi-periodic boundary condition 
$x_0 (t) =0.5+0.4\sin (\omega t)$, with 
$\omega =\pi(\sqrt{5}-1)$. These results can be 
summarized by saying that, basically, there exist four 
qualitatively different spatio-temporal behaviors.
\begin{description} 

\item[A)] Non chaotic and convectively stable. The comoving
Lyapunov exponents are negative for all values of $v$ 
($\lambda (v) < 0$, $\forall v$) and $x_n \to 0$ for 
$n \to \infty$. The region of the parameter space corresponding 
to this behavior (absolute stability) is identified by the 
`$\Box$' symbols in Fig. \ref{fig:4_16}: we call it ``region $A$''. 
One can say that the quasi-periodic (or chaotic) boundary condition 
$x_0 (t)$ is not able to excite the bulk of the system 
(see Fig. \ref{fig:4_1}).  

\item[B)] Non chaotic and marginally convectively stable. The 
comoving Lyapunov exponents have a maximum value equal to zero 
for a $v^{\ast} \neq 0$ : $\lambda_{max}(v)=\lambda(v^{\ast})=0$
(the region with `$+$' symbols, in  Fig. \ref{fig:4_16}, that 
we call ``region $B$''). The quasi-periodic boundary 
conditions produce spatio-temporal ``strips patterns'' 
(see Fig. \ref{fig:4_3}). 

\item[C)] Non chaotic but convectively unstable. The maximal 
Lyapunov exponent, $\lambda_1 = \lambda (0)$, is negative, but 
the comoving Lyapunov exponent spectrum is positive in a 
certain interval of $v > 0$ (the region with `$\ast$' symbols, 
in  Fig. \ref{fig:4_16}, that we call ``region $C$''). The 
spatio-temporal behavior appears irregular (see Fig. \ref{fig:4_13}).
   
\item[D)] Standard chaotic (the region with `\rule{2mm}{2mm}',
in Fig. \ref{fig:4_16}, that we call ``region $D$''). In this 
case one has $\lambda (0) > 0$, and the spatio-temporal 
behavior is irregular, and is similar to the one of Fig. 
\ref{fig:4_13}.
\end{description}

We remark that the results discussed above remain valid, 
with slight changes, for chaotic boundary conditions;
e.g. with $x_0 (t)$ given by the $y$ variable of the H\'enon 
map: $y(t+1)=-\alpha y(t)^2+\beta y(t-1)+1$, with typical values 
of the parameters $\alpha =1.4$ and $\beta =0.3$.

\section{Quantitative characterization of the spatial 
         behavior}
We discuss here how to characterize the convectively 
unstable region $C$ of Fig. \ref{fig:4_16}. Part of the 
results in this section have been briefly discussed in ref. 
\cite{new}. Some authors, e.g. Pikovsky \cite{piko} and Kozlov 
{\it et al.} \cite{koz}, stressed the fact that the 
``irregularity'' of these systems 
seems to increase with $n$. An analysis of $x_n$ as a function 
of $t$ (by means of some standard methods for the 
characterization of dynamical systems, like, for instance, 
the power spectrum, the return map, the Grassberger-Procaccia 
correlation dimension \cite{gras}) typically shows that $x_1$ 
is more irregular than $x_0$, $x_2$ more irregular than $x_1$, 
and so on. Fig. \ref{fig:3_4} shows $x_n (t+1)$ vs $x_n (t)$ 
for different $n$: it is evident an increasing of the irregularity 
as $n$ increases. A simple way to characterize quantitatively 
the spatial complexity is by studying the spatial correlation 
functions: 
\begin{equation}
\label{eq:corre}
C (n, m) = {\langle x_n x_{m} \rangle - 
\langle x_n \rangle \langle x_{m} \rangle \over \langle x^2 _n 
\rangle - \langle x_n \rangle ^2} , 
\end{equation}
where the average is with respect to the time. $C (n, m)$ vs $m$, 
computed in the convectively unstable region, is shown in 
Fig. \ref{fig:3_5}, for different $n$; two facts are evident:
\begin{itemize} 
\item the shape of $C (n, m)$, at least for $n \gg 1$, 
does not depend on $n$, but only on $|n-m|$, thus revealing 
that we are observing a bulk property of the system;  
\item the correlation decays exponentially, 
      $C (n, m)\sim \exp (-|n-m|/\xi)$.
\end{itemize}  

It is natural to wonder how an uncertainty $\delta x_0 (t)= 
O(\epsilon)$ -- with $\epsilon \ll 1$ -- on the knowledge of the 
boundary conditions will affect the system. In this paper we 
consider only the case of infinitesimal perturbations, so 
that we may safely assume that $\delta x_n$ evolves according 
to the tangent vector equations of the system (\ref{eq:uno}):
\begin{equation}
\label{eq:unobis}
\delta x_n(t+1) = (1-c) f_{a}' (x_n(t)) \delta  x_n(t) 
             + c  f_{a}' (x_{n-1}(t)) \delta  x_{n-1}(t) .
\end{equation}  
For the moment we do not consider, for sake of simplicity,
intermittency effects, that is, we neglect finite time 
fluctuations of the comoving Lyapunov exponents.  The 
uncertainty $\delta x_n (t)$, on the determination of 
the variable at the site $n$, is given by the superposition 
of the evolved $\delta x_0 (t-\tau)$ with $\tau =n/v$:
\begin{equation}
\label{eq:due}
\delta x_n(t) \sim \int \delta x_0 (t-\tau) e ^{\lambda(v)\tau} dv
     = \epsilon \int e ^{ [\lambda(v)/v] n} dv. 
\end{equation}
Since we are interested in the asymptotic spatial behavior, 
i.e. the large $n$ one, we can write:
\begin{equation}
\label{eq:tre}
\delta x_n(t)\sim \epsilon e ^{\gamma n},
\end{equation}
where, in the particular case of a non intermittent 
system, one has:
\begin{equation}
\label{eq:quattro}
\gamma=\max _v \frac {\lambda(v)}{v}. 
\end{equation}
Equation  (\ref{eq:quattro}) is a link between the 
comoving Lyapunov exponent and the ``spatial''  Lyapunov 
exponent $\gamma$, a more precise and operative 
definition of which is given by:
\begin{equation}
\label{eq:cinque}
\gamma = \lim _{n \to \infty} \frac {1}{n}\Biggl
\langle \ln \frac {|\delta x_n|} {\epsilon} \Biggr \rangle , 
\end{equation}
where the brackets mean a time average.

So equation (\ref{eq:quattro}) establishes a relation between 
the convective instability of a system and its sensitivity 
to the boundary conditions, which can be considered a sort of 
spatial complexity. 

We remark again that equation (\ref{eq:quattro}) holds 
exactly only in the absence of intermittency; 
in this case it can be shown from (\ref{eq:quattro}) that 
our spatial index can be written in a simple way in terms of 
another kind of spatial Lyapunov exponents, $\mu (\Lambda)$, 
introduced in ref. \cite{torc}:
\begin{equation}
\label{eq:torcin}
\gamma=\max _v \frac {\lambda(v)}{v}= \mu(\Lambda =0) . 
\end{equation}

In the general case the relation is rather more 
complicated. We introduce the effective comoving 
Lyapunov exponent, $\tilde \lambda _t (v)$, that gives 
the exponential changing rate of a perturbation, in the 
frame of reference moving with velocity $v$, on a finite 
time interval $t$. 

Then, instead of (\ref{eq:due}) we obtain 
\begin{equation}
\label{eq:delta}
\delta x_n(t) \sim \epsilon 
     \int e ^{[\tilde \lambda _t (v)/v] n} dv,  
\end{equation}
and therefore:
\begin{equation}
\label{eq:sei}
\gamma = \lim _{n \to \infty} \frac {1}{n}\Biggl
\langle \ln \frac {|\delta x_n|} {\epsilon} \Biggr \rangle =
\lim _{n \to \infty} \frac {1}{n} 
\ln \frac {|\delta x_n^{typical}|} {\epsilon}=\Biggl
\langle \max _v \frac {\tilde \lambda _t (v)} {v} \Biggr \rangle .
\end{equation}
In a generic case, because of the fluctuations, it is not 
possible to write $\gamma$ in terms of $\lambda (v)$.
Nevertheless it is possible to state a lower bound:
\begin{equation}
\label{eq:sette}
\gamma \geq \max _v 
\frac {\langle \tilde \lambda _t (v) \rangle} {v}=
\max _v \frac {\lambda (v)} {v}\equiv \gamma^{\ast}.
\end{equation}
The evaluation of the function $\lambda (v)$ needs a 
heavy computational effort, however one can find good 
approximations of the quantity $\gamma^{\ast}$. A first 
simple approximation for it, actually a lower bound, is 
given by 
\begin{equation}
\label{eq:otto}
\gamma_1 = \frac {\lambda(v^{\ast})}{v^{\ast}} , 
\end{equation}
where $v^{\ast}$ is the velocity at which $\lambda$
attains its maximum value. The analysis of the long time 
behavior of many impulsive perturbations makes it 
possible to obtain $v^{\ast}$ and $\lambda(v^{\ast})$ 
without the knowledge of $\lambda (v)$ as a function of $v$. 
An improvement of this approximation can be performed in 
the following way. Beside $\lambda(v^{\ast})$, one computes
the usual Lyapunov exponent $\lambda_1 =\lambda (0)$, then 
one estimates the function $\lambda(v)$, by assuming it is 
the parabola $\lambda_p (v)$ passing in the point $(0, \lambda_1)$ 
with maximum $\lambda(v^{\ast})$ for $v=v^{\ast}$ and, 
finally, one determines $\gamma_p = \max _v  [\lambda _p (v)/v]$. 
Typically $\gamma_p$ is very close (within a few percent) to 
$\gamma^{\ast} $. 

In Fig. \ref{fig:1old} we show $\gamma$, $\gamma^{\ast}$, 
$\gamma_1$ and $\gamma_p$ {\it versus} $a$ at a fixed value 
of $c$ ($c=0.7$), again for the logistic map with the
quasi-periodic boundary condition $x_0 (t)
=0.5+0.4\sin ((\sqrt{5}-1)\pi t)$.
There is a large range of values of the parameter $a$ for 
which $\gamma$ is rather far from $\gamma^{\ast}$; for instance, 
at $a=3.74$ we have $\gamma=0.28$ and $\gamma ^{\ast}=0.26$. 

The difference is an effect of the intermittency; this may be 
pointed out by looking at what happens with the map $f_a (x)=ax$ 
mod $1$: in this case we find that, all over the explored range 
of variation of $a$, $\gamma$ and $\gamma^{\ast}$, from a 
numerical point of view, are indistinguishable (their relative 
difference is smaller than $10^{-6}$). 

We may obtain a further indication of the fact that the non 
negligible fluctuations of the comoving Lyapunov exponents 
are at the origin of the marked difference of $\gamma$ from 
its lower bound, by introducing, following ref. \cite{ben}, 
the generalized spatial Lyapunov exponents, $L_s (q)$. 
These quantities allow us to characterize the fluctuations 
in the growth of the perturbations along the chain:  
\begin{equation}
\label{eq:nove}
L_s (q) = \lim _{n \to \infty} \frac {1}{n}
\ln \Biggl \langle \Biggl\vert \frac {\delta x_n} 
{\epsilon}\Biggr\vert^q \Biggr \rangle . 
\end{equation}
By means of standard arguments of probability theory, 
one has that: 
\begin{itemize} 
\item $L_s (q)/q$ is a monotonic non decreasing function 
of $q$;
\item $dL_s(q)/dq|_{q=0}=\gamma$;
\item $L_s (q) = \gamma \, q + \frac {1}{2} \sigma ^2 q^2 $,
  for  small $q$, where $\sigma ^2 = \lim _{n \to \infty}
 \langle \bigl( \ln \bigl|\delta x_n / \epsilon \bigr|
  - \gamma n\bigr)^2 \rangle /n$ . 
\end{itemize}
The shape of $L_s (q)/q$ depends on the details of the 
dynamics, however $L_s (q)$ is fairly determined -- 
at least for small values of $q$ -- by the two parameters 
$\gamma$ and $\sigma ^2$. The reason for having introduced 
this function is that one expects some relation between the 
fluctuations of the spatial-complexity index $\gamma$, and 
the fluctuations of the effective comoving Lyapunov exponents, 
and it is much more easy to compute the former than the latter. 
Fig. \ref{fig:2old}, shows that, in the case of the logistic 
map, as we expected, the parameter $\sigma ^2$ (that is related to 
the variance of the spatial fluctuations) is small (large) in 
the region where $\gamma ^{\ast}$ is a good (bad) approximation 
of $\gamma$. 

We stress that the definition (\ref{eq:cinque}) and 
the bound (\ref{eq:sette}) have a general validity. 
It is not difficult to understand that they are 
valid not only for 1-D flow systems, such as model (\ref{eq:uno}), 
but also for continuous-time systems (as in the case of the 
asymmetric Ginsburg-Landau equation). As a matter of fact, all 
the arguments discussed above hold unaltered whenever one can 
introduce the comoving Lyapunov exponents.

At the end of this section we want to note that there 
is not a simple relation between the correlation length 
$\xi$ and the exponent $\gamma$, such as, for instance, 
$\xi \propto \gamma ^{-1}$. Indeed -- in analogy with the 
case for the corresponding quantities (characteristic 
correlation time and maximal Lyapunov exponent) used to 
characterize the temporal behavior of the dynamical 
systems with few degrees of freedom -- we do not expect 
a simple relation between $\xi$ and $\gamma$.

\section{Transition to chaos}
From Fig. \ref{fig:4_16} one can see that, for the
system under investigation, there exist two routes
for reaching the chaos:
\begin{description} 
\item[I)] the way from non chaotic and convectively 
stable behavior, i.e. from region $A$ to region $D$;
\item[II)] the way from non chaotic, but convectively 
unstable behavior, i.e. from region $C$ to region $D$.
\end{description}
In this section we study with a certain detail this twofold
way to chaos, by looking at the features of the Lyapunov 
exponents of the system. 

It is easy to understand that, for the system (\ref{eq:uno})
the computation of all the Lyapunov exponents is much 
easier than in the generic case. The origin of this lucky 
fact is in the triangular structure of the Jacobian 
matrix ${\bf M} [{\bf x}(t)]$ that rules the evolution
of the tangent vector
\begin{equation}
\label{eq:dodici}
\delta {\bf x} (t+1) =  {\bf M} [{\bf x}(t)] 
            \delta {\bf x} (t) ,
\end{equation}  
where $\delta {\bf x}=(\delta x_1,\delta x_2, \dots, 
\delta x_N)$, ${\bf x}= (x_1, x_2, \dots, x_N)$ and 
\begin{equation}
\label{eq:tredici}
{\bf M} [{\bf x}]=  \pmatrix{
                   {(1-c)f_{a}'(x_1)}&{0}&{0}&{0}&{\ldots}\cr
                   {cf_{a}'(x_1)}&{(1-c)f_{a}'(x_2)}&{0}&{0}&{\ldots}\cr
                   {0}&{cf_{a}'(x_2)}&{(1-c)f_{a}'(x_3)}&{0}&{\ldots}\cr
                   {\vdots}&{\vdots}&{\vdots}&{\vdots}&{\ddots}\cr
                      } .
\end{equation}
Since the product of triangular matrices is again a triangular 
matrix, all the Lyapunov exponents $\lambda _1, \lambda _2, 
\dots, \lambda _N$ (ordered, as usual, according to the decreasing 
values) can be computed in a simple way -- without using the 
standard orthonormalization method of Benettin {\it et al.} 
\cite{benet} -- from the quantities  
\begin{equation}
\label{eq:quattordici}
\Lambda _i = \ln (1-c) + \lim _{t \to \infty}
{1 \over t} \sum _{n=1} ^{t} \ln |f' _a (x_i (n))|, 
\qquad i=1,2,\dots, N.
\end{equation}
The Lyapunov exponents $\{ \lambda _j \}$ then are nothing 
but the $\{ \Lambda _i \}$ after a reordering by
decreasing values. We stress that, for the system under 
investigation, the computer time $T_N$ for the computation 
of all the Lyapunov exponents is proportional to the 
system size $N$: $T_N \sim N$; while in a generic map, 
with local coupling, one has $T_N \sim N^2 .$ 

The transition of type ${\bf I})$ -- that takes place in 
the zone of the region $D$ close to the boundary with the 
region $A$ -- shows the behavior already observed in the
maps with a generic symmetric local coupling \cite{W}. By 
looking at Fig. \ref{fig:4_18} one clearly sees that there 
exists a very well established thermodynamic limit for the 
Lyapunov spectrum. When $N\to\infty$ there exists a limiting 
function $G(x)$ such that 
\begin{equation}
\label{eq:quindici}
\lambda _i \simeq G(i/N)  .
\end{equation}
The existence of this limit entails various consequences.
From 
(\ref{eq:quindici}) one can conclude that the number 
of the non negative Lyapunov exponents $N_0$ is proportional 
to $N$:
\begin{equation}
\label{eq:10sette}
N_0 \sim N  .
\end{equation}
So, by the Pesin formula
\cite{eck}, one obtains a finite Kolmogorov-Sinai entropy 
per degree of freedom, $h$:
\begin{equation}
\label{eq:sedici}
h=\lim _{N\to \infty} {H\over N} =\lim _{N\to \infty} 
{1\over N} \sum _{i=1}^{N} \lambda _i \theta (\lambda _i) =
\int _0 ^1 G(x)\theta (G(x)) dx , 
\end{equation}
where $H$ is the Kolmogorov-Sinai entropy and $\theta$ is the 
step function. In addition, from eq. (\ref{eq:quindici}) and the 
Kaplan-Yorke conjecture \cite{eck} one infers that the information 
dimension $d_I$ of the attractor is proportional to $N$:
\begin{equation}
\label{eq:10otto}
d_I \sim N  .
\end{equation}

For the transition of type ${\bf II})$ -- taking place in the 
part of the region $D$ close to the boundary with the region 
$C$ -- the features of the Lyapunov exponents are very different 
from those described above. Fig. \ref{fig:4_21a} shows that the 
behavior of the $\{ \lambda _i \}$ do not follow eq. 
(\ref{eq:quindici}); on the contrary one has 
\begin{equation}
\label{eq:10nove}
\lambda _i \simeq F(i)  .
\end{equation}
Therefore, in this case, the Kolmogorov-Sinai entropy and 
the information dimension of the attractor are not extensive
quantities: $H= O(1)$ and $d_I = O(1) \quad \forall N.$  
Loosely speaking we can say that this transition can be 
described in terms of a finite layer.

At the end of this section we want to stress the following 
point. In Fig. \ref{fig:4_21a}, as a matter of fact, we show 
$\Lambda _i$ vs $i$ since, in the case of the transition 
of type ${\bf II})$, the ordering of $\Lambda _i$ according 
to their decreasing values coincides with their ordering 
according to the label of the lattice site on which they are 
computed (see eq.(\ref{eq:quattordici})): $\Lambda _n \geq 
\Lambda _{n+1}$ (with the exception of very few sites close to 
the boundary); this means that the horizontal coordinate in Fig. 
\ref{fig:4_21a} is just the site label of the chain. 
Therefore one realizes that, in this case, the positive Lyapunov 
exponents are in correspondence with the sites close to the 
boundary: this gives further support to the idea that we are 
observing a kind of finite layer phenomenon. On the other hand, 
for the transition of type ${\bf I})$, where one has a good 
thermodynamic limit, there is no correspondence between 
the sites on the lattice and the Lyapunov exponents. 
This is well evident from Fig. \ref{fig:4_19a}. 

\section{Conclusions and Discussion}
In this paper we characterized, in a quantitative way, 
the spatial complex behavior and the transition to chaos 
of flow systems. We have shown that in a non chaotic, but 
convectively unstable flow -- where the convective 
instability induces a spatial sensitivity to the boundary 
conditions -- it is possible to introduce an index (a sort 
of ``spatial'' Lyapunov exponent) for the quantitative 
characterization of this ``spatial complexity''. Moreover, 
there exists a relation (a bound) between this spatial 
complexity and the comoving Lyapunov exponents. 

The transition to chaos can take place in two possible 
scenarios: either from a state that is both non chaotic and 
spatially non complex, or from a state that is non chaotic but 
convectively unstable. In the first case one has a standard 
thermodynamic limit: the Kolmogorov-Sinai entropy and the 
information dimension of the attractor are proportional to 
the system size. In the second case one has a sort of layer 
phenomenon, where only a finite number of Lyapunov exponents 
-- related to the sites near the boundary -- are positive. 

We conclude noting that all the results above do not depend 
too much on the details of the used boundary conditions 
$x_0 (t)$. Indeed we have that if $x_0 (t)$ has a chaotic 
behavior (like, for instance, that of a variable of the 
H\'enon map) $\gamma$ and $\sigma ^2$, as functions of 
$a$ are not very different from the case with $x_0 (t)$ a 
quasi-periodic function. The same is true for the properties 
of the two kind of transitions to chaos.
 
\acknowledgments
We thank K. Kaneko and A. Pikovsky for useful discussions and
correspondences. We thank A. Politi, S. Ruffo and A. Torcini 
for having pointed out to us the existence of eq. (\ref{eq:torcin}).

\begin{figure}
\caption{The behavior of the system (\ref{eq:uno}) in the space 
of the control parameters $c$ and $a$: `$\Box$' = absolute 
stability; `$+$'= marginal convective stability; `$\ast$'= convective 
instability; `\protect\rule{2mm}{2mm}' = absolute instability.}
\label{fig:4_16}
\end{figure}

\begin{figure}
\caption{Evolution of a state of the system, with $a=2.94$ and 
    $c=0.7$ (region $A$); the boundary condition is quasi-periodic: 
    $x_0 (t) = 0.5 + 0.4 \sin (\omega t)$, with 
    $\omega =\pi(\sqrt{5}-1)$.}
\label{fig:4_1}
\end{figure}

\begin{figure}
\caption{Evolution of a state of the system, with $a=3.40$ and 
      $c=0.7$ (region $B$); the boundary condition is quasi-periodic, 
      as in Fig. \ref{fig:4_1}.  For a better graphical 
      effect only the configurations at even times 
      have been reported.}
\label{fig:4_3}
\end{figure}

\begin{figure}
\caption{Evolution of a state of the system, with $a=3.85$ and 
      $c=0.7$ (region $C$); the boundary condition is quasi-periodic, as in  
      Fig. \ref{fig:4_1}. }
\label{fig:4_13}
\end{figure}

\begin{figure}
\caption{Graph of $[x_n (t), x_n (t+1)]$ at different $n$; 
        the values of the parameters are $c=0.7$, $a=3.9$ 
       (region $C$); the boundary condition is quasi-periodic, as in  
        Fig. \ref{fig:4_1}. }
\label{fig:3_4}
\end{figure}

\begin{figure}
\caption{Spatial correlation function $C (n, m) = 
  [\langle x_n x_{m} \rangle - \langle x_n \rangle 
   \langle x_{m} \rangle] / [\langle x^2 _n \rangle - 
   \langle x_n \rangle ^2]$, as a function of $m$, 
    for $n=100$ (a), $n=200$ (b) and $n=400$ (c)
    with parameters $c=0.7$, $a=3.85$ (region $C$); the boundary 
    condition is quasi-periodic, as in Fig. \ref{fig:4_1}. }
\label{fig:3_5}
\end{figure}

\begin{figure}
\caption{$\gamma$ ($+$), $\gamma_1$ ($\times$), $\gamma_p$ ($\Box$) 
 and $\gamma ^{\ast}$ ($\bigcirc$) vs $a$ 
     at fixed $c=0.7$; the boundary condition is quasi-periodic, as in  
      Fig. \ref{fig:4_1}.}
\label{fig:1old}
\end{figure}

\begin{figure}
\caption{$\sigma ^2$ ($+$) and $\gamma - \gamma ^{\ast}$ ($\Box$) 
        vs $a$ at $c= 0.7$; the boundary condition is 
        quasi-periodic, as in Fig. \ref{fig:4_1}. }
\label{fig:2old}
\end{figure}

\begin{figure}
\caption{$\lambda _i$ vs $i$ for the system (\ref{eq:uno}), 
        with $c=0.06$ and $a=3.58$, in the cases  
       $N=100$ (dotted line), $N=200$ (dashed line) 
        and $N=400$ (full line); the boundary condition is 
        quasi-periodic, as in Fig. \ref{fig:4_1}.}
\label{fig:4_18}
\end{figure}

\begin{figure}
\caption{$\lambda _i$ vs $i$ for the system (\ref{eq:uno}), 
        with $c=0.27$ and $a=3.70$, in the cases  
       $N=200$ (full line) and $N=400$ (dotted line);
       the boundary condition is quasi-periodic, as in  
       Fig. \ref{fig:4_1}. }
\label{fig:4_21a}
\end{figure}

\begin{figure}
\caption{Non ordered Lyapunov spectra $\Lambda _i$ vs $i$ 
         for the system (\ref{eq:uno}), with $N=200$ 
         $c=0.06$ and $a=3.58$, 
         obtained with two different initial conditions 
         (identified with `$\circ$' or `$\star$')}
\label{fig:4_19a}
\end{figure}

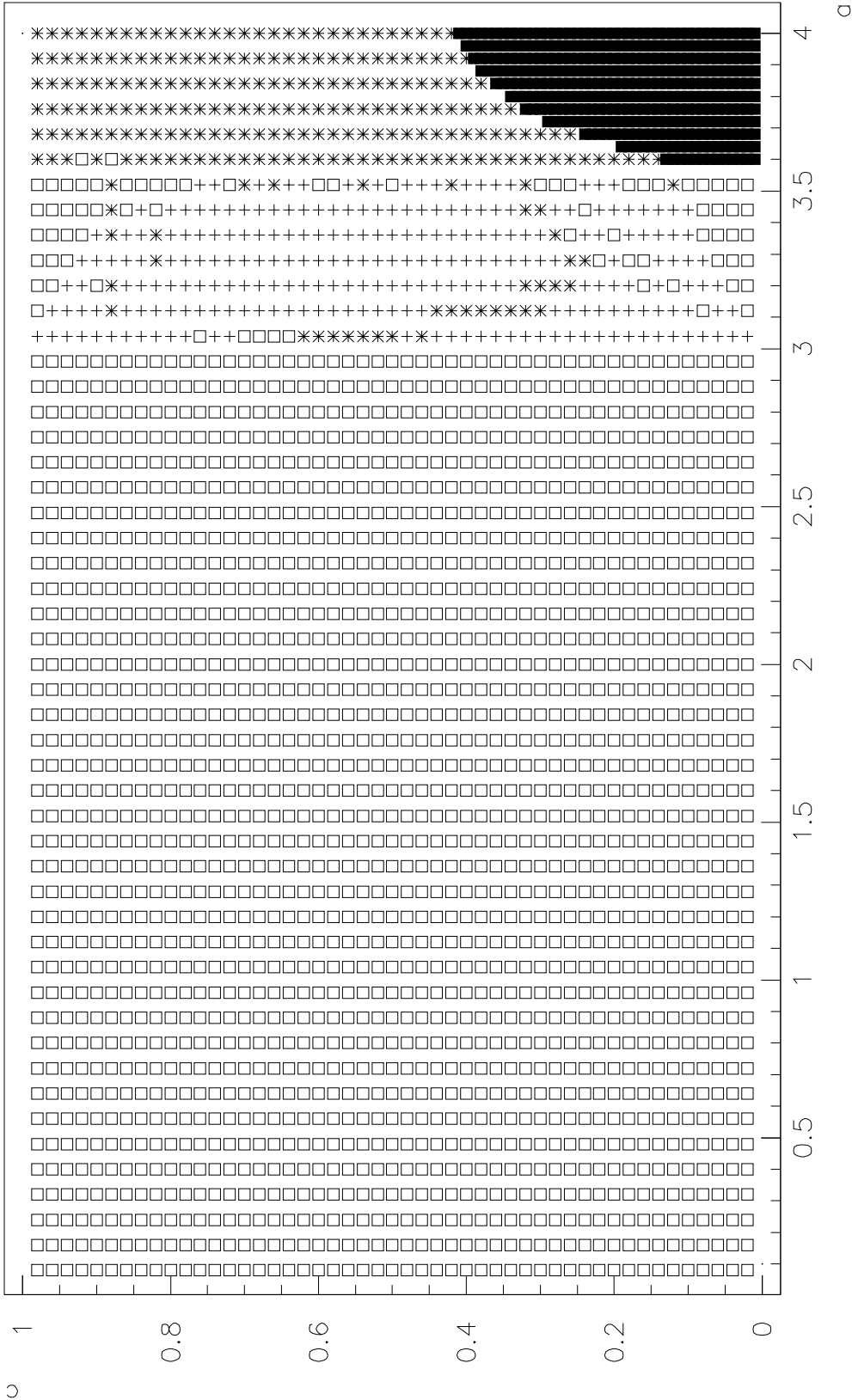
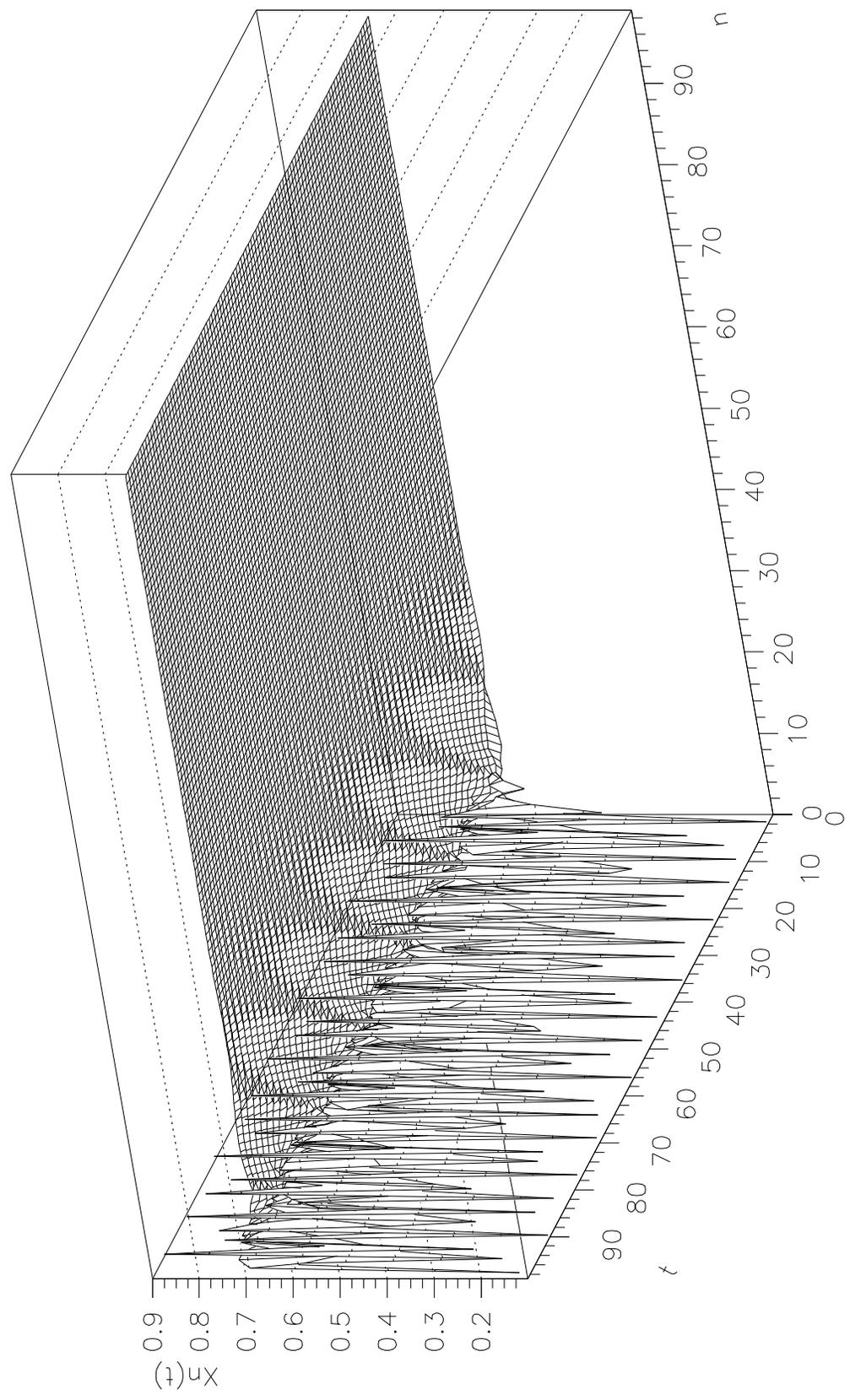
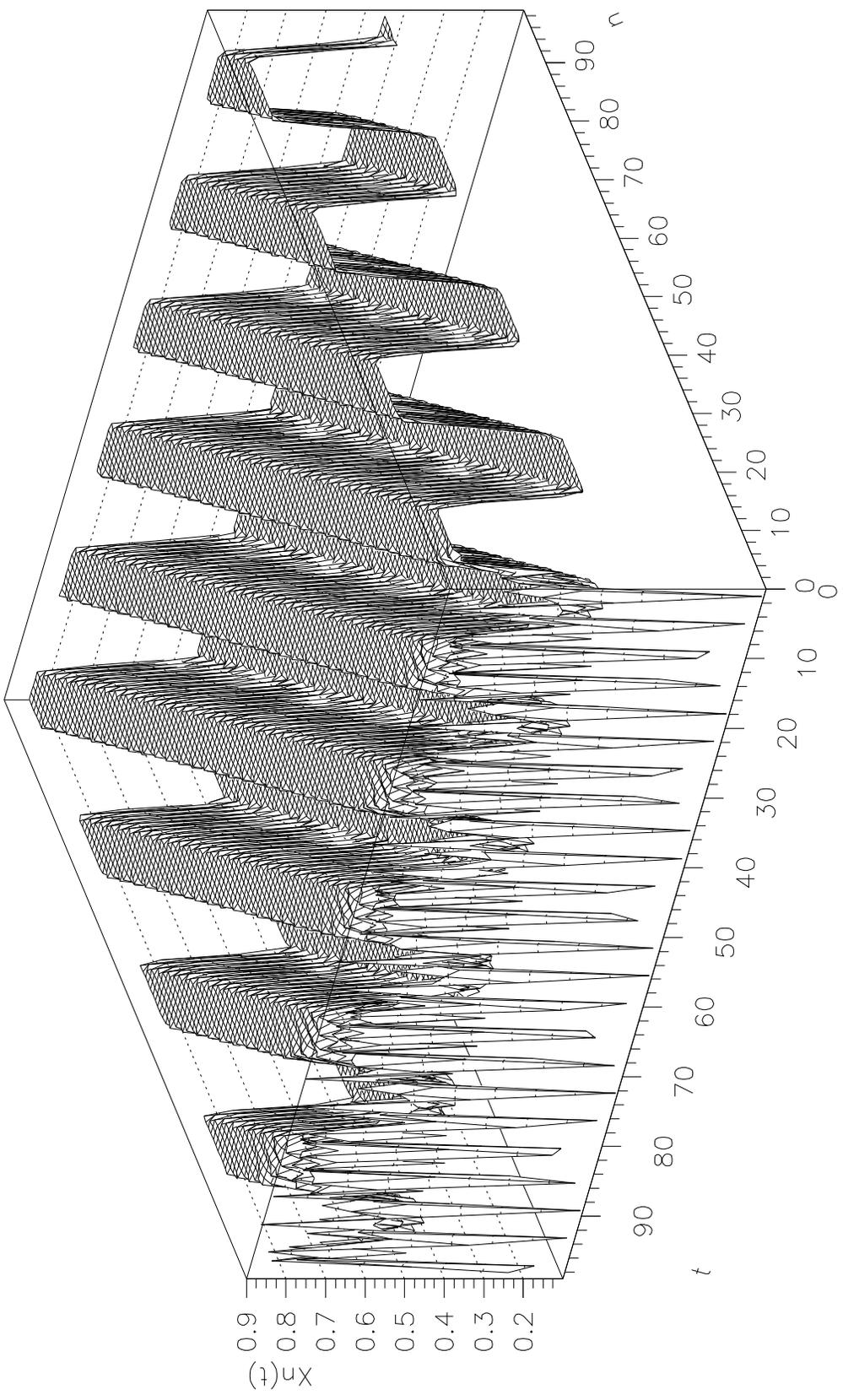
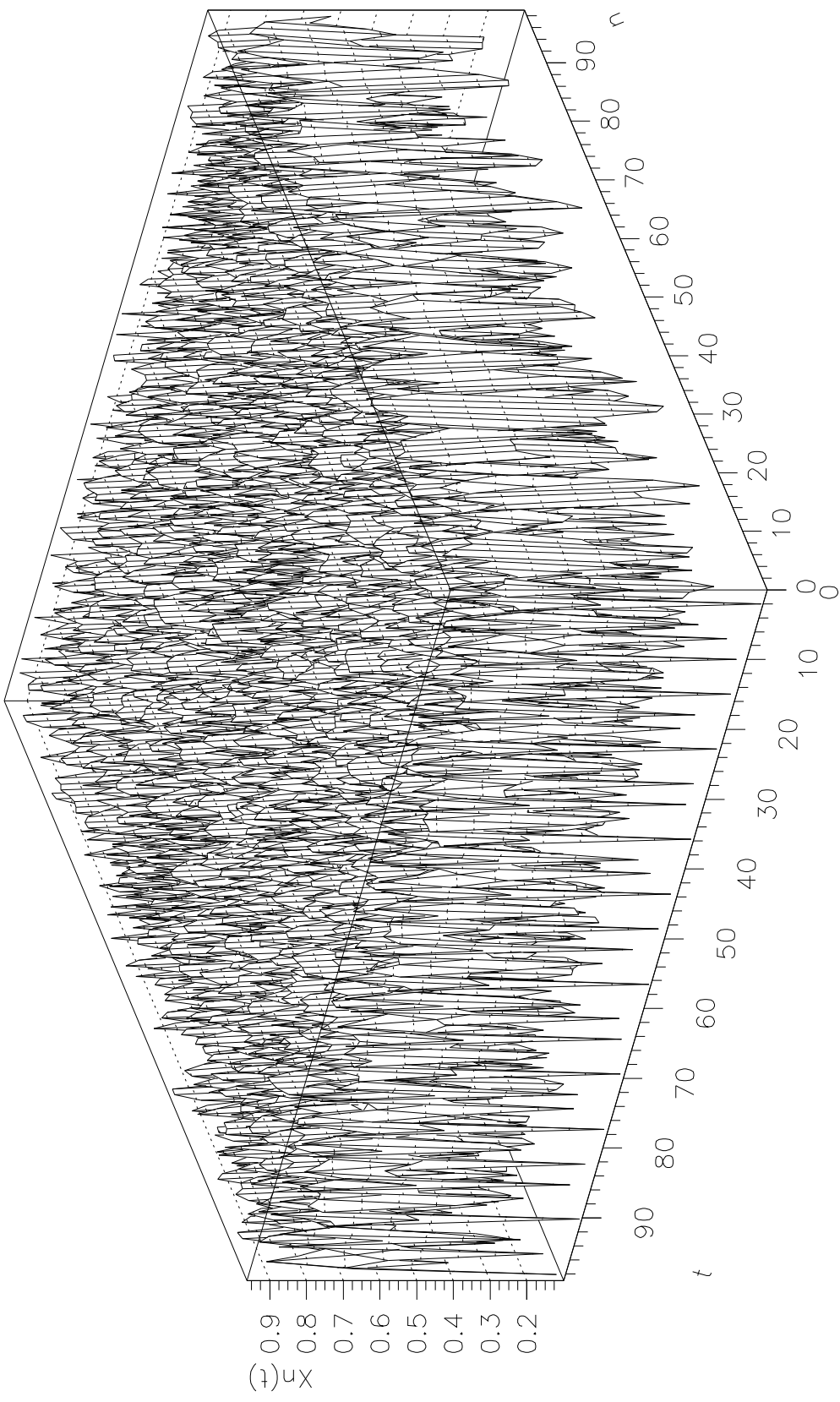
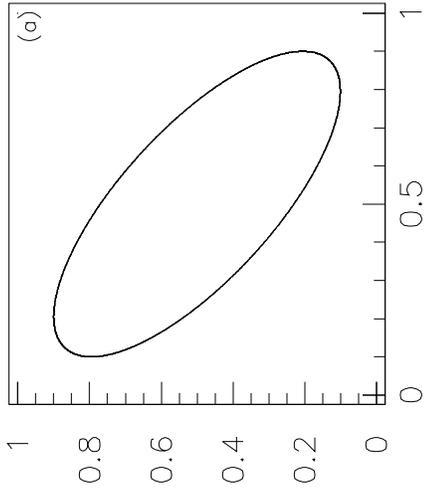
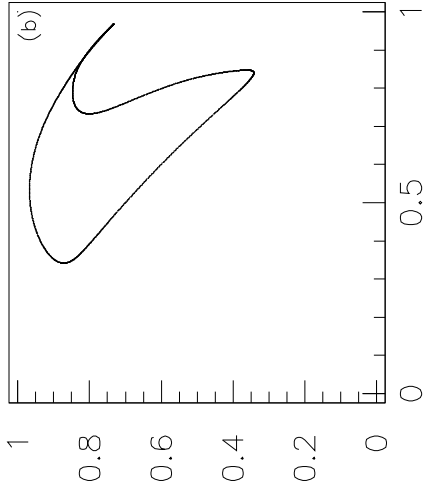
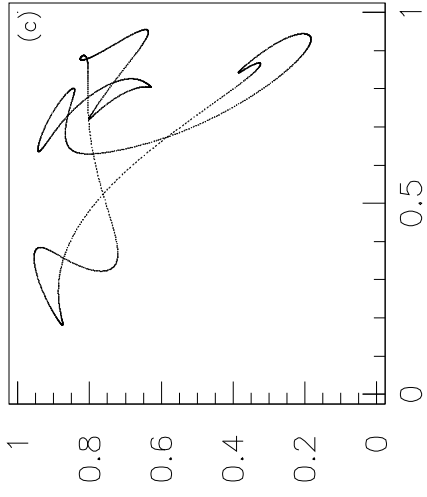
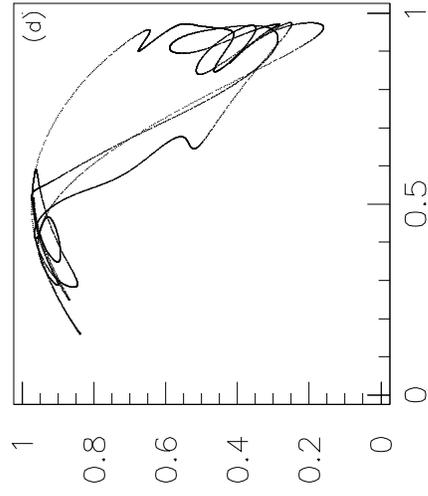
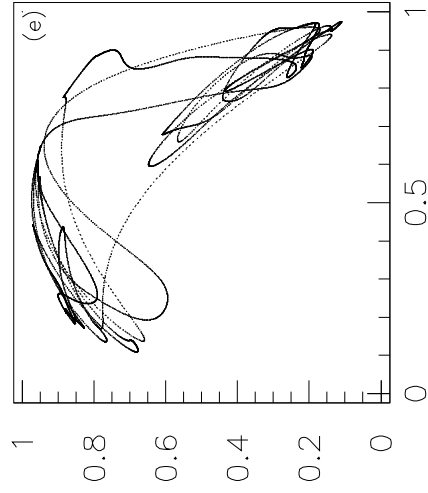
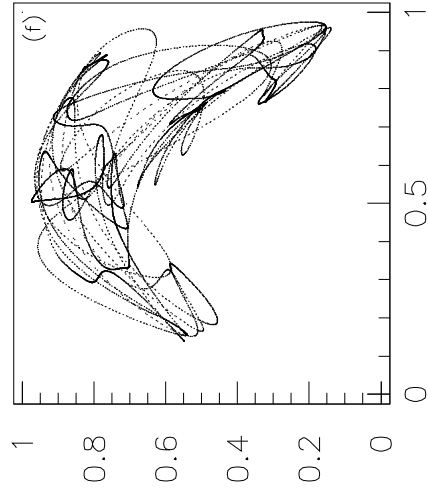
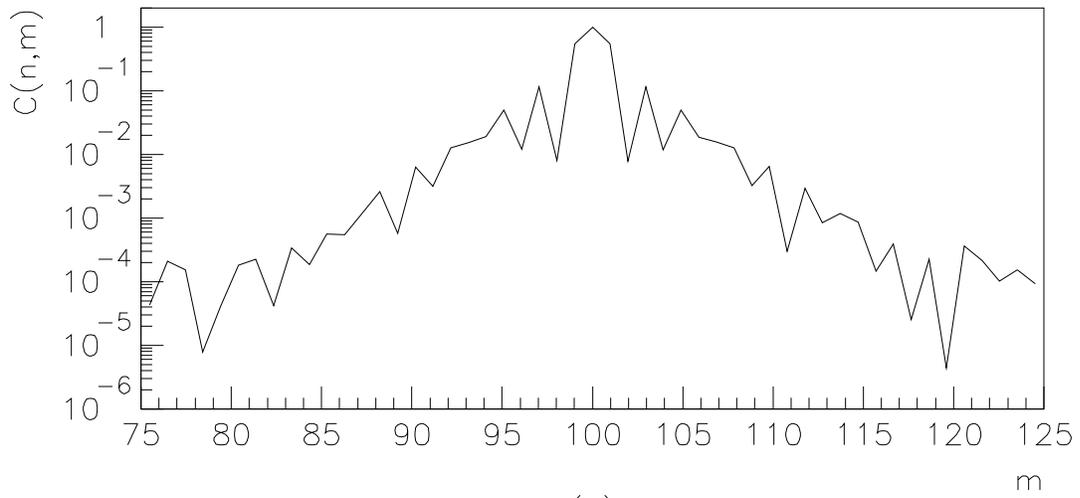
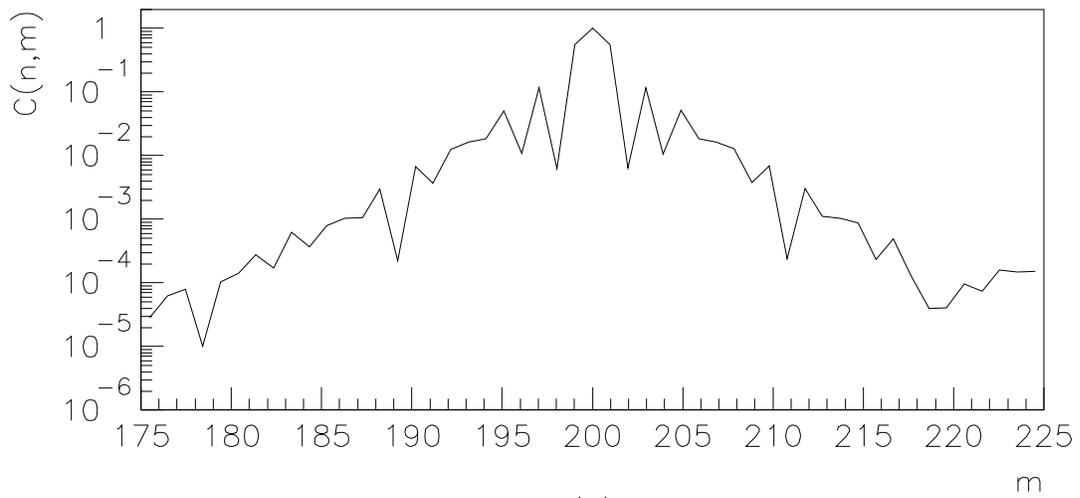
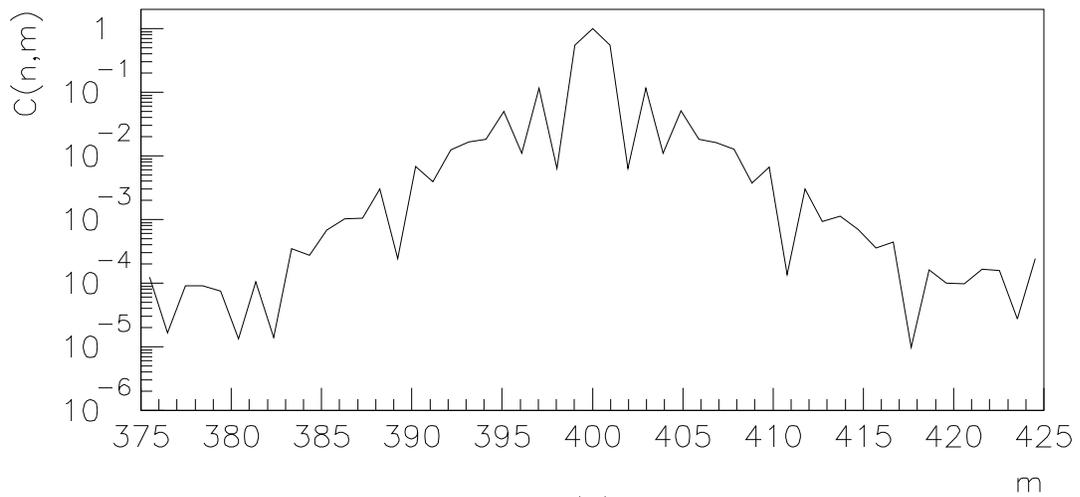
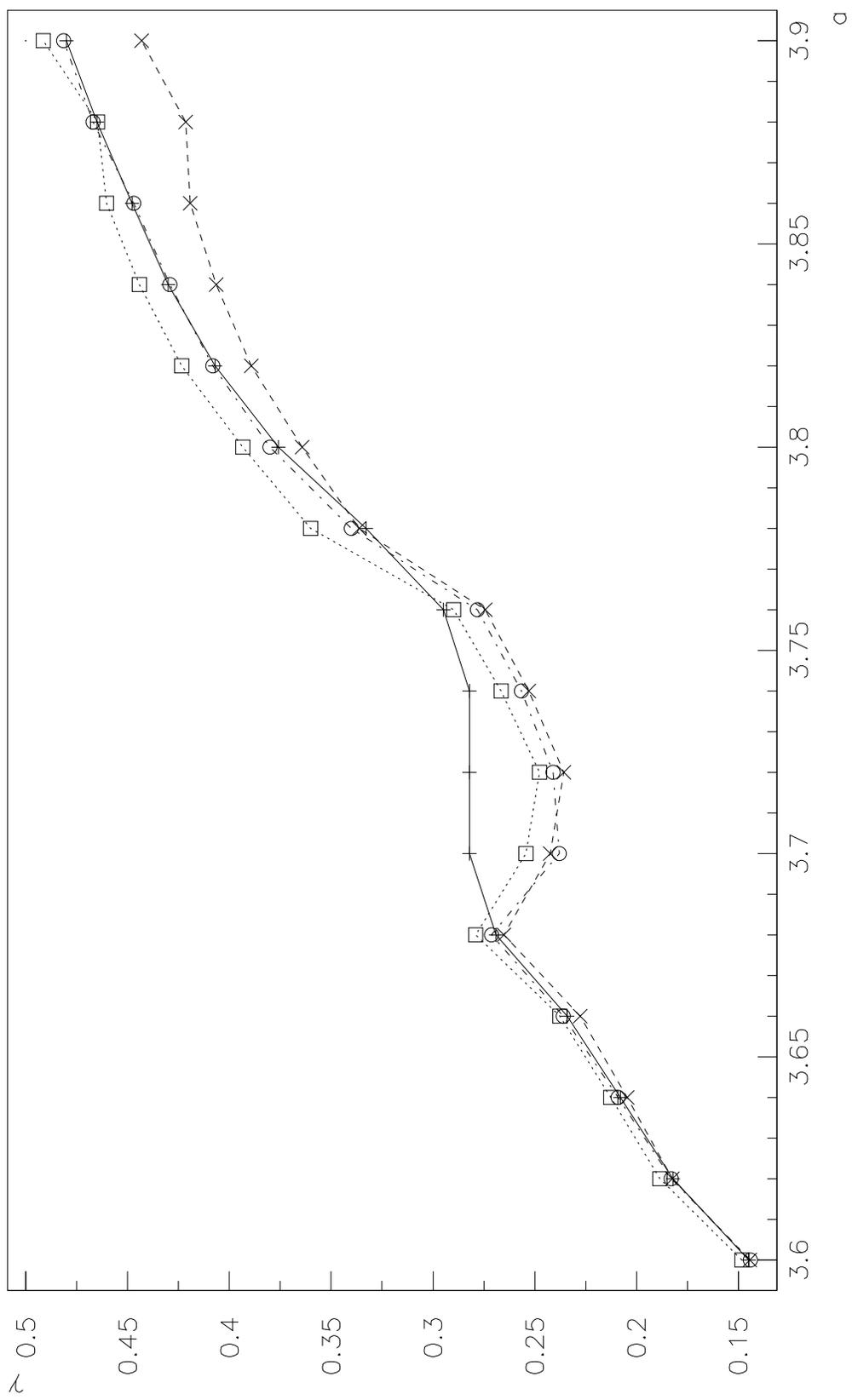
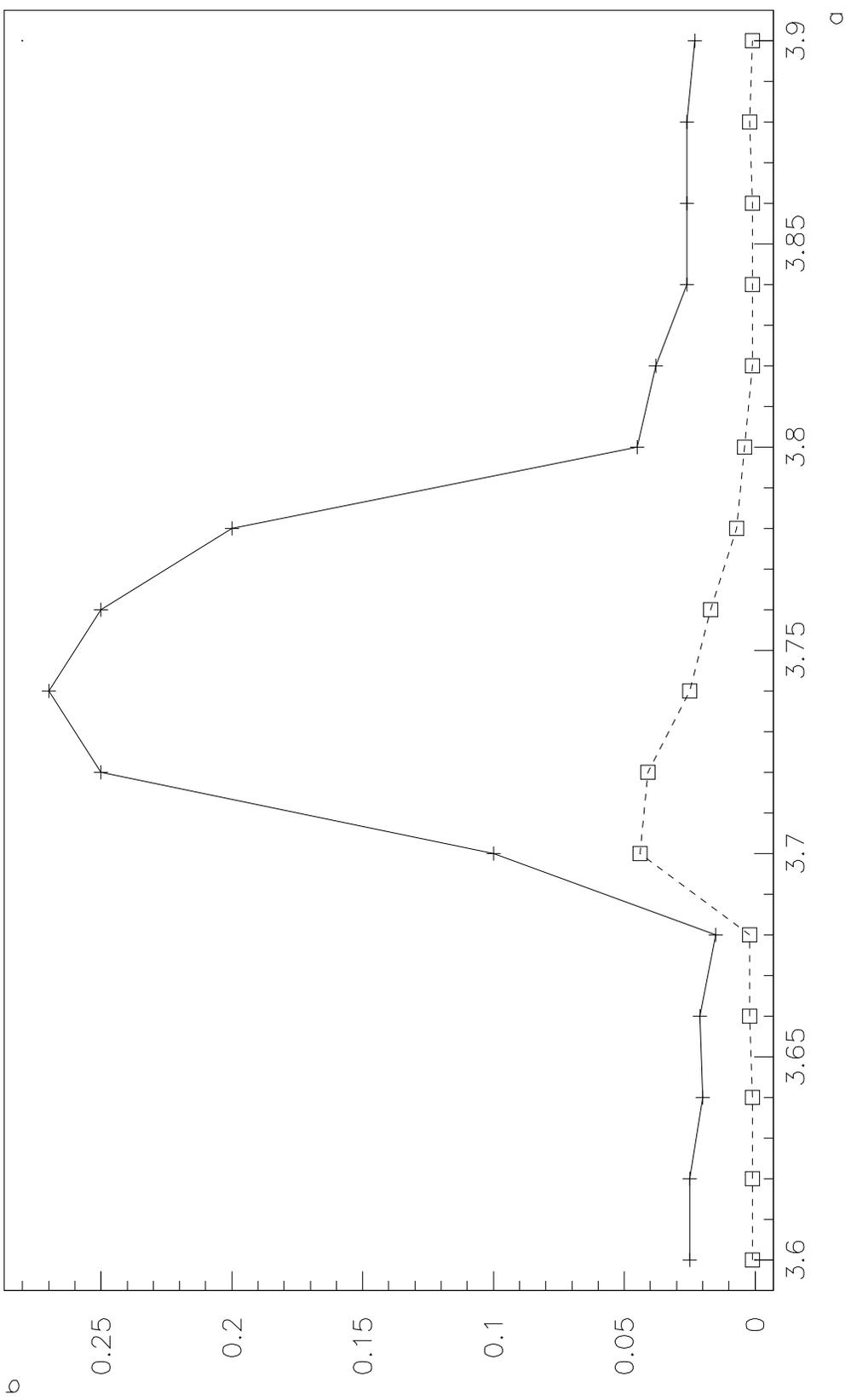
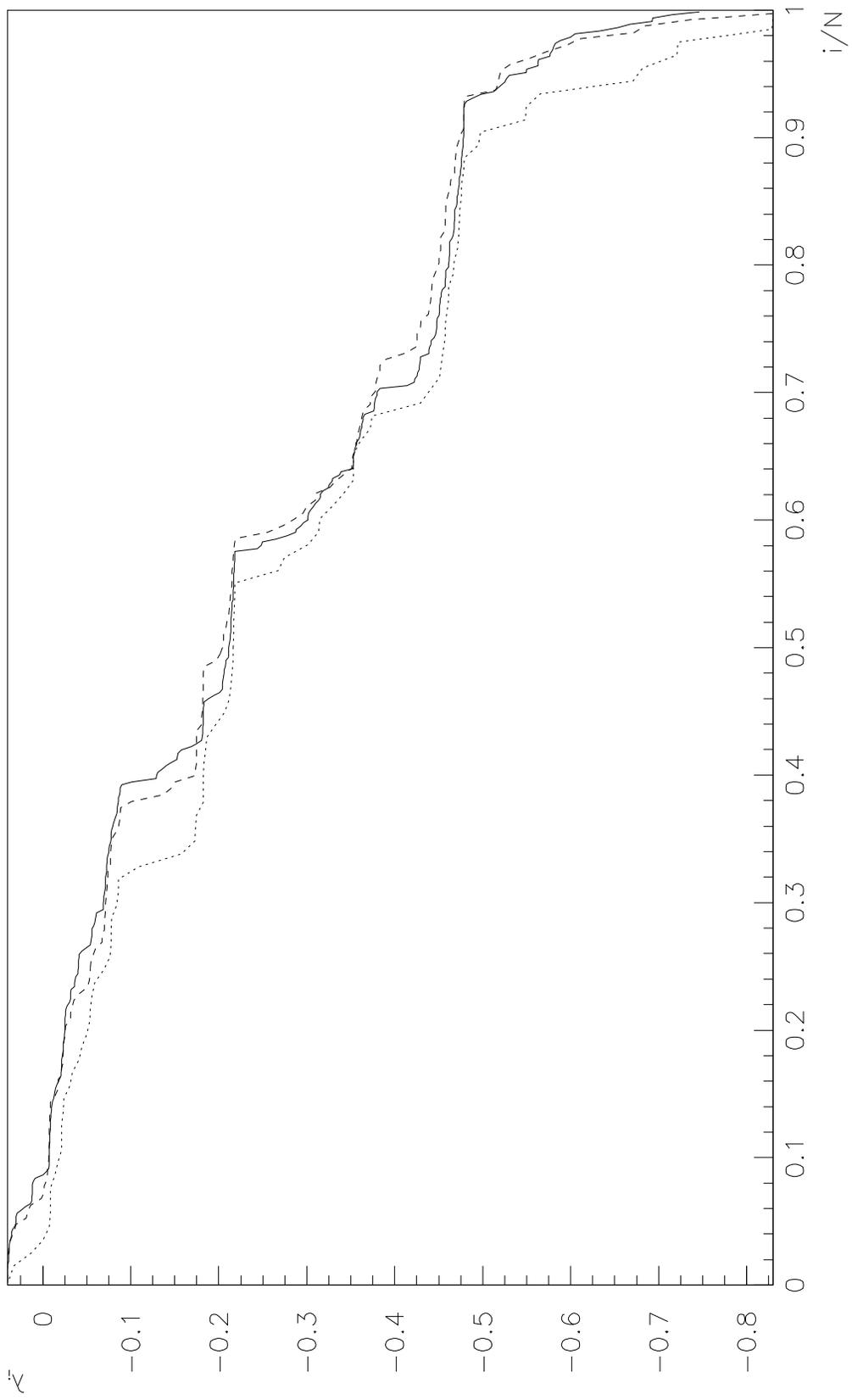
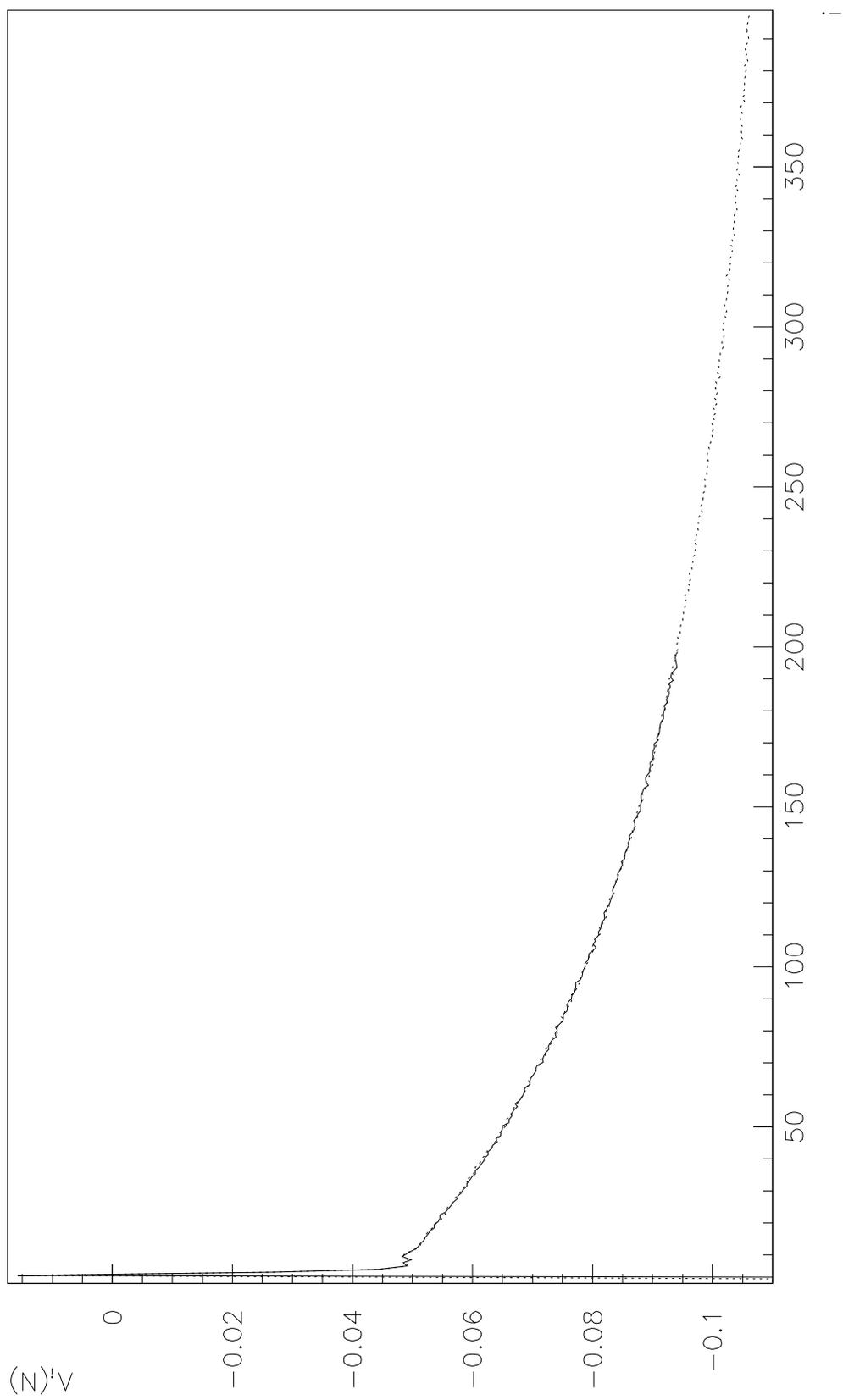
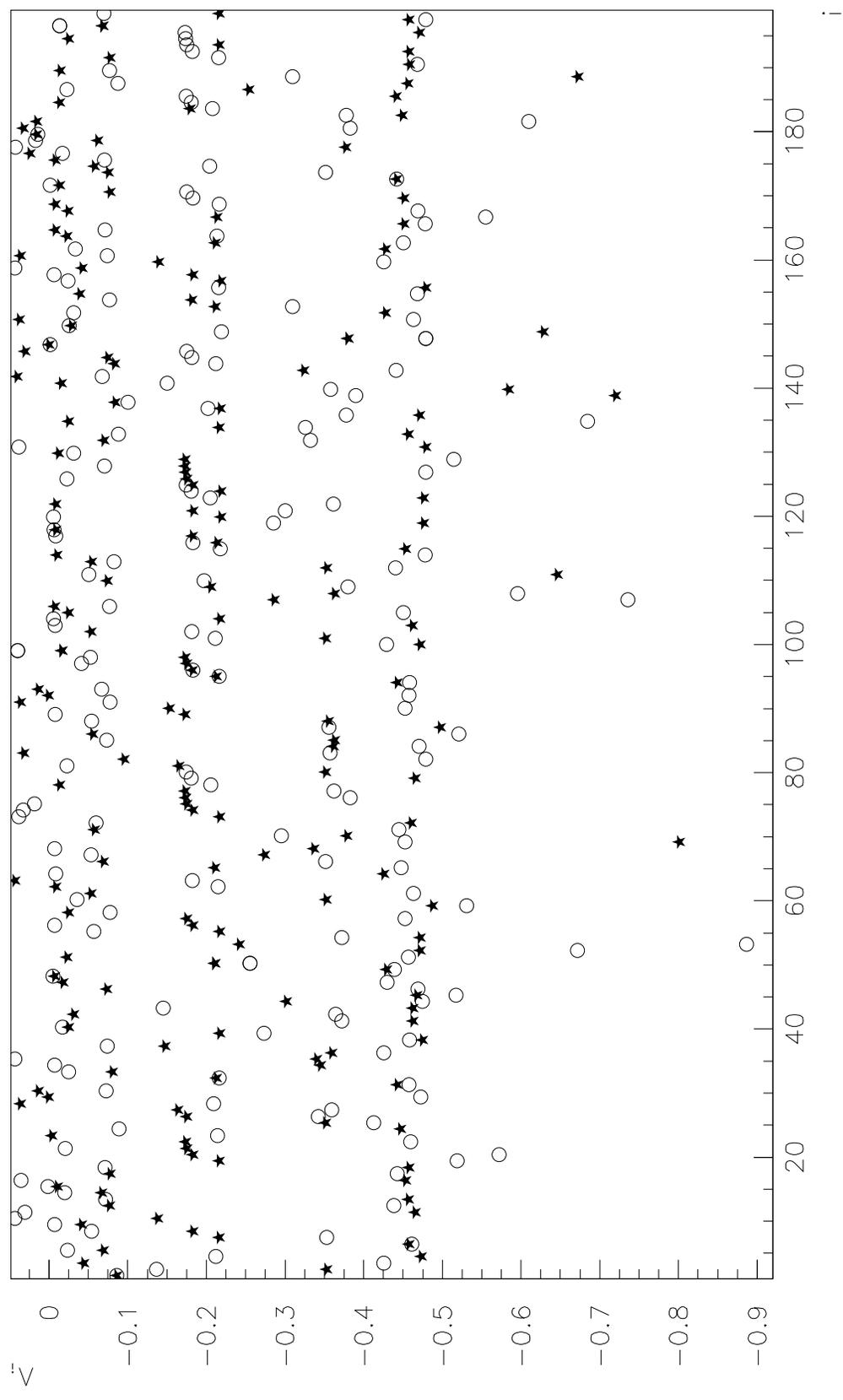
\end{document}